\documentclass[onecolumn,preprintnumbers,prb,fleqn]{revtex4}
\usepackage{graphicx,color}
\usepackage{amsmath,amssymb}
\newcommand{\be}{\begin{equation}} 
\newcommand{\ee}{\end{equation}} 
\newcommand{\bea}{\begin{eqnarray}}
\newcommand{\eea}{\end{eqnarray}}

\begin{document}
\title{Tight binding parameters for graphene}
\author{Rupali Kundu}
\email{rupali@iopb.res.in}
\affiliation{ Institute of Physics, Bhubaneswar 751005 India.\\}
\date{\today}
\begin{abstract}
In this article we have reproduced the tight binding $\pi$ band dispersion of graphene including upto third nearest neighbours and also calculated the partial density of states (due to $\pi$ band only) within the same model. The aim was to find out a set of parameters descending in order as distance towards third nearest neighbour increases compared to that of first and second nearest neighbours with respect to an atom at the origin. Here we have talked about two such sets of parameters by comparing the results with first principle band structure calculation \cite{1}.
\end{abstract}
\maketitle
\section{Introduction}
Graphene is a single sheet of atomic thickness with carbon atoms arranged hexagonally. Though it was an ideal two dimensional material of theoretical interest and one of the earliest material on which tight binding band structure calculation was done\cite{2,3}, it has triggered recently a lot of interest among people including reinvestigation of many earlier results since its experimental discovery in 2004\cite{4}. Particularly, a large no of people have recalculated the tight binding band with nearest neighbour hopping but without overlap integral correction\cite{1,2,3,5,6,7,8,9}, some have calculated the same by taking into account the overlap integral correction\cite{2, 8}, out of these only few calculations are there which take care of second and third nearest neighbours along with overlap integral corrections\cite{1,9}. It is noticed that the first nearest neighbour hopping integral ($\gamma_0$) lies around 2.5eV-3.0eV when tight binding band is fitted with first principle calculation or experimental data\cite{1,6,8} near the $K$ point of the brillouin zone of graphene but interestingly, when one tries to have a good matching of the tight binding band over the whole brillouin zone by including upto third nearest neighbour hoppings and overlap integrals, the tight binding parameters are considered as merely fitting parameters, not as physical entities\cite{1} i.e, the values of parameters do not decrease consistently as one moves towards second and third nearest neighbours. In present work we have fitted our tight binding band with first principle data with an objective to get a set of parameters which is free from the above discrepancy and found out a set which gives good matching with the first principle data over the whole brillouin zone. Here we have calculated the density of states also with second and third nearest neighbours. The article has been arranged in the following manner: section I talks about the geometrical structure of graphene; the electronic structure and density of states (DOS) of graphene with nearest, next nearest and next to next nearest neighbour couplings are discussed in section II and finally, section III summarizes and concludes over the previous sections.
\section{Geometrical structure of graphene}
Since the geometrical structure of a material plays a crucial role in determining the electronic dispersion of the material, it is important to look at the details of the structure of graphene before going into the discussion of band structure of graphene. The structure of an ideal graphene sheet is a regular hexagonal arrangement of carbon atoms in two dimension as shown in fig. 1. It consists of two inequivalent (with respect to orientations of bonds) triangular sublattices called A-sublattice and B-sublattice. The unit cell contains one A and one B type of carbon atoms contributed by respective sublattices. Each carbon atom has three nearest neighbours coming from the other sublattice, six next nearest neighbours from the same sublattice and three next to next nearest neighbours from the other sublattice. $a_{0}$ (1.42\AA) is the nearest neighbour lattice distance. In the figure, $\vec a_{1}$ and $\vec a_{2}$ are the unit vectors with magnitude $a=\sqrt{3}a_{0}$ i.e. 2.46\AA. With respect to $A_{0}$ atom the coordinates of the first neighbours ($B_{1i}, i=1,2,3$); second neighbours ($A_{2i}, i=1,...6$) and third neighbours ($B_{3i}, i=1,2,3$) are $\left( a/\sqrt{3},0\right), \left(-a/2\sqrt{3}, -a/2\right), \left(-a/2\sqrt{3}, a/2\right)$; $\left(0, a\right), \left(\sqrt{3}a/2, a/2\right), \left(\sqrt{3}a/2, -a/2\right), \left(0, -a\right), \left(-\sqrt{3}a/2, -a/2\right), \left(-\sqrt{3}a/2, a/2\right)$ and $\left(a/\sqrt{3}, a\right), \left(a/\sqrt{3}, -a\right), \\
\left(-2a/\sqrt{3}, 0\right)$ respectively.
{\begin{figure}[h]
       \centering 
       \includegraphics[width=8cm]{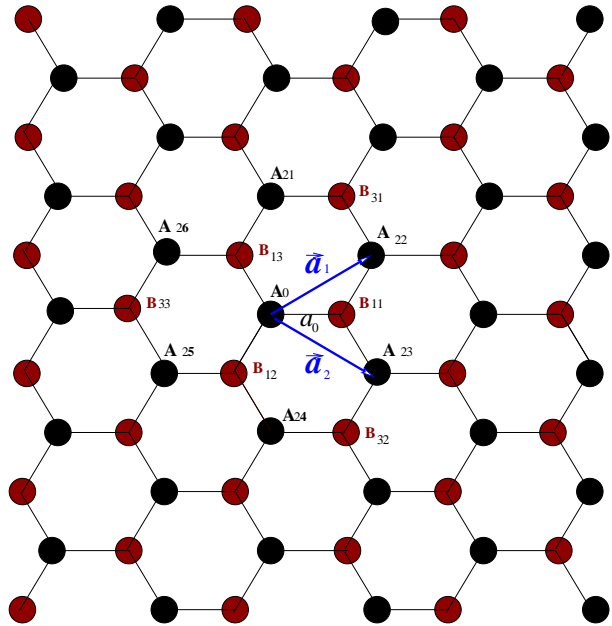}
		\caption{Structure of graphene. The structure with black circles forms the A-sublattice and that with red circles gives B-sublattice. $\vec a_1$ and $\vec a_2$ are the unit vectors.}
	\end{figure}}
\section{Electronic structure of graphene}
Carbon atom has six electrons with the electronic configuration $1s^{2}2s^{2}2p^{2}$. 2s and 2p levels of carbon atoms can mix up with each other and give rise to various hybridized orbitals depending on the proportionality of s and p orbitals. Graphene has $sp^{2}$ hybridization: 2s orbital overlaps with $2p_{x}$ and $2p_{y}$ orbitals and generates three new inplane $sp^{2}$ orbitals each having one electron. The ${2p_z}$ orbital remains unaltered and becomes singly occupied. Due to overlap of $sp^{2}$ orbitals of adjacent carbon atoms strong $\sigma$ (bonding) and $\sigma^{*}$ (antibonding) bonds are formed. The bonding $\sigma$ bonds, lying in a plane, make an angle of $120^{\circ}$ with each other and is at the root of hexagonal planar structure of graphene. $p_{z}$ orbitals being perpendicular to the plane overlap in a sidewise fashion and give $\pi$ (bonding) and $\pi^{*}$ (antibonding) bonds. $sp^{2}$ orbitals with a lower binding energy compared to $1s$ (core level) are designated as semi core levels and $p_{z}$ orbitals having lowest binding energy are the valence levels. Overlapping of $p_z$ energy levels gives the valence band (bonding $\pi$ band) and conduction band (antibonding $\pi^*$ band) in graphene. Thus, we see that while the structure of graphene owes to $\sigma$ bonds, $\pi$ band is responsible for the electronic properties of graphene and hence as far as electronic properties of graphene are concerned, concentration is given only on $\pi$ bands. Since the $p_z$ orbitals overlap in a sidewise manner, the corresponding coupling is weaker compared to that of $\sigma$ bonds (where $sp^{2}$ orbitals overlap face to face). So the $p_{z}$ orbitals almost retain their atomic character. Hence, to describe the electronic structure of graphene, tight binding model could be a good choice.
\subsection{Tight binding band}
In this subsection tight binding bands of graphene have been reproduced including upto third nearest neighbor hopping of electrons and overlap integral corrections with focus on the point to find out tight binding parameters which will not look like just fitting parameters rather look somewhat like physical parameters. The results have been compared with existing literature\cite{1}. The key equations are shown here for the different cases of nearest neighbours, next nearest neighbours and next to next nearest neighbours and the details of the calculations are given in the appendix. Since there are two atoms per unit cell coming from two sublattices, the total wave function can be written as
\be
\Psi_{k}(r)=C_{A}\Psi_{A}^{k}+C_{B}\Psi_{B}^{k},
\ee
where $\Psi_{A}^{k}(r)=1/\sqrt{N}\Sigma_{A}e^{i\vec{k}.\vec{r_{A}}} \Phi_{A}(r-r_{A})$ and $\Psi_{B}^{k}(r)=1/\sqrt{N}\Sigma_{B}e^{i\vec{k}.\vec{r_{B}}} \Phi_{B}(r-r_{B})$ are the tight binding Bloch wave functions from A and B sublattices. Here, N is the number of unit cells in the crystal, $C_{A}$ and $C_{B}$ are contributions coming from A and B sublattices respectively, $\Phi^{'}s$ are $2p_{z}$ atomic orbitals, $k$ is crystal momentum and $r_{A}$ and $r_{B}$ are the positions of A and B atoms respectively with respect to a chosen origin.
If $H$ is the Hamiltonian and $E(k)$ the eigenvalue then
\be
H\Psi_{k}(r)=E(k)\Psi_{k}(r),
\ee
which leads to the secular equation
\be
\left|\begin{array}{cc}
\ H_{AA} - E(k)S_{AA} & H_{AB}-E(k)S_{AB}       \\
 H^*_{AB}-E(k)S^*_{AB} & H_{BB}-E(k)S_{AA}     \\
\end{array}\right|=0
\ee
Since two sublattices are equivalent, $H_{AA}=H_{BB}$ and $S_{AA}=S_{BB}$ and the general dispersion relation follows as 
\be
E^{\pm}(k)=\left[(2E_0-E_1)\pm \sqrt{(E_1-2E_0)^2-4E_1E_2}\right]/2E_3 \text{,}
\ee
where $S_{AA}H_{AA}=E_0$, $H_{AB}S_{AB}^*+H^*_{AB}S_{AB}=E_1$, $H_{AA}^2-H_{AB}H^*_{AB}=E_2$ and $S_{AA}^2-S_{AB}S^*_{AB}=E_3$.
The explicit forms of the dispersions under various situations are discussed below.\\
\underline{\bfseries{With Nearest Neighbour Approximation:}}
In case of nearest neighbour approximation contribution comes from nearest atoms of the other sublattice. Detailed calculation of all the matrix elements is shown in appendix. From equation (4) the dispersion relation for this case becomes
\be
E^\pm(k)=\left[ \left( E_{2p}-s_0\gamma_0g(k)\right)\pm \left( \gamma_0-s_0E_{2p}\right) \sqrt{g(k)}\right] /\left[ 1-s^2_0 g(k)\right],
\ee
where $E_2p$, $\gamma_0$ and $s_0$ are site energy, nearest neighbour hopping and overlap integrals respectively.\\
\underline{\bfseries{With Second Nearest Neighbour Approximation:}}
In this part the modification in the energy spectrum due to the presence of second nearest neighbor atoms will be discussed. The matrix elements $H_{AA}$ and $S_{AA}$ get modified but $H_{AB}$ and $S_{AB}$ remain as they were. The corresponding energy momentum relation is
\be
E^\pm(k)=\left[ E_{2p}+\gamma_{1}u(k)\mp \gamma_{0}\sqrt{g(k)}\right]/\left[ 1+s_{1}u(k)\mp s_{0}\sqrt{g(k)}\right],
\ee 
where $\gamma_1$ and $s_1$ are next nearest neighbour hopping and overlap integrals respectively.\\
\underline{\bfseries{With Third Nearest Neighbor Approximation:}}
In this case the matrix elements $H_{AB}$ and $S_{AB}$ get changed but $H_{AA}$ and $S_{AA}$ remain unaltered.
Calculation of all the matrix elements gives expressions for $E_1$, $E_2$, $E_3$ as follows:
\bea
E_1&=& 2s_0\gamma_0 g(k)+\left(s_0 \gamma_2+\gamma_0 s_2\right) t(k)+2s_2\gamma_2 g(2k)\nonumber \\
E_2&=&\left[E_{2p}+\gamma_1u(k)\right]^2-\left[ \gamma^{2}_0g(k)+\gamma_0 \gamma_2 t(k) +\gamma^{2}_2 g(2k)\right]\nonumber\\
E_3&=&\left[1+s_1u(k)\right]^2-\left[s^{2}_0 g(k)+ s_0 s_2 t(k) + s^{2}_2 g(2k)\right],\nonumber
\eea
which when put in equation (4) gives the energy momentum relation in this case. $\gamma_2$ and $s_2$ being next to next nearest neighbour hopping and overlap integrals respectively. The bands of graphene for above three cases are plotted along the high symmetry directions of its hexagonal brillouin zone sketched in figure (2).
 {\begin{figure}[h]
       \centering
       \includegraphics[width=6.0cm]{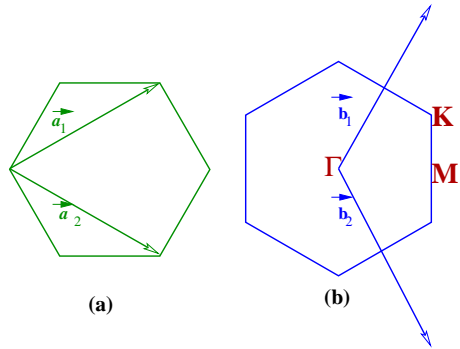}
		\caption{Brillouin zone of graphene. (a) Real space unit vectors. (b) k-space unit vectors and high symmetry directions. }
	\end{figure}}
As shown in the figure, the symmetry points are $\Gamma(0,0)$, $M(2\pi/\sqrt{3}a, 0)$ and $K(2\pi/\sqrt{3}a, 2\pi/3a)$. There are six corner points, out of which three are independent since the nearby corner points($K$ and $K^{\prime}$) are inequivalent.
In fig. (3) we have compared the bands with nearest, next nearest and next to next nearest neighbour hopping and overlaps with a first principle calculation [produced from this model with the parameters $E_2p=-0.36$eV, $\gamma_0=-2.78$eV, $\gamma_1=-0.12$eV, $\gamma_2=-0.068$eV, $s_0=0.106$, $s_1=0.001$, $s_2=0.003$ because within the chosen energy scale it does not show any energy difference with first principle data\cite{1}] for the set of parameters shown in table I where we have first determined values of nearest neighbour parameters ($\gamma_0$ and $s_0$) which best reproduces the first principle result, then we have taken care of second nearest neighbours ($\gamma_1$ and $s_1$) keeping first nearest neighbour parameters fixed and lastly considered the third nearest neighbour parameters ($\gamma_2$ and $s_2$) with fixed ($\gamma_0$, $s_0$) and ($\gamma_1$, $s_1$) with the expectation to have better matching over the whole brillouin zone of graphene.
\begin{figure}[h]
       \includegraphics[width=7cm]{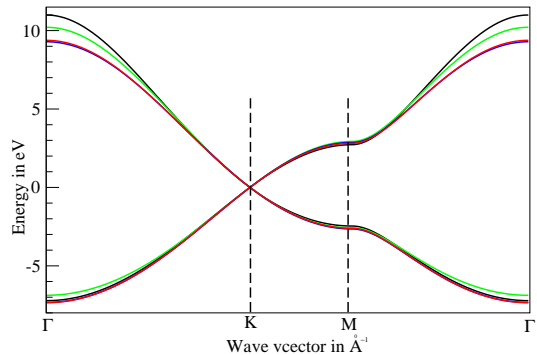}
        	\caption{Electronic structure of graphene with first principle result (black curve), first nearest neighbour interactions (green curve), second nearest neighbour interactions (blue curve) when first nearest neighbour parameters are fixed and third nearest neighbour interactions (red curve) when first and second nearest neighbour parameters are fixed. The parameters are listed in Table I.}
	\end{figure} 
\begin{table}[h]
\caption{Tight binding parameters}
\begin{center}
\begin{tabular}{|c|r|r|r|r|r|r|r|} 
\hline\hline
Neighbours & $E_2p$(eV) & $\gamma_0$(eV) & $\gamma_1$(eV) & $\gamma_2$(eV) & $s_0$ & $s_1$ & $s_2$  \\
\hline
1st & 0.0 & -2.74 &  & & 0.065  &   &   \\
\hline
2nd & -0.21 & -2.74 & -0.07 &   &0.065  & 0.002  &   \\
\hline
3rd & -0.21 &  -2.74 & -0.07 & -0.015 & 0.065 & 0.002 & 0.001 \\
\hline 
\end{tabular}
\end{center}
\end{table} 
We observe that nearest neighbour coupling gives good overall matching for both valence and conduction bands but with the inclusion of second and third nearest neighbours under the above restrictions, the overall matching of the valence band with the first principle band improves whereas for the conduction band it is good in the optical range. When only nearest neighbour hopping is considered, the total band width (difference between valence band and conduction band energies) at $\Gamma$ point is $6|\gamma_0|$, that at $M$ point is $2|\gamma_0|$ but when nearest neighbour overlap integral is included the valence and conduction band energies at $\Gamma$ point occur at $3\gamma_0/(1+3s_0)$ and $-3\gamma_0/(1-3s_0)$ respectively and those at $M$ point appear at $\gamma_0/(1+s_0)$ and $-\gamma_0/(1-s_0)$ respectively. The $K$ point energy is zero for both the cases. When next nearest neighbours are included the energy at $K$ point is $(E_{2p}-3\gamma_1)/(1-3s_1)$. So the values of $E_{2p}$ and $\gamma_1$ are properly chosen to have $K$ point energy at zero. In this case the valence and conduction band energies at $\Gamma$ point are at $(E_{2p}+6s_1+3\gamma_0)/(1+6s_1+3s_0)$ and $(E_{2p}+6s_1-3\gamma_0)/(1+6s_1-3s_0)$ and those at $M$ point are at $(E_{2p}+2s_1+\gamma_0)/(1+2s_1+s_0)$ and $(E_{2p}+2s_1-\gamma_0)/(1+2s_1-s_0)$ respectively. In fig. (4) we have plotted the above set of curves with a different set of parameters such that both nearest neighbour and next nearest neighbour parameters are free when the effect of second nearest neighbour is looked for and all three, nearest, next nearest and next to next nearest neighbour parameters are free when the third nearest neighbours are included. The parameters are shown in table II. This figure shows that inclusion of second nearest neighbour gives better result over first nearest neighbour interaction compared to the plots in fig. 3. Consideration of third nearest neighbour couplings gives very good matching for both valence and conduction bands along all the high symmetry directions of the brillouin zone.
 {\begin{figure}[h]
       \centering
       \includegraphics[width=7cm]{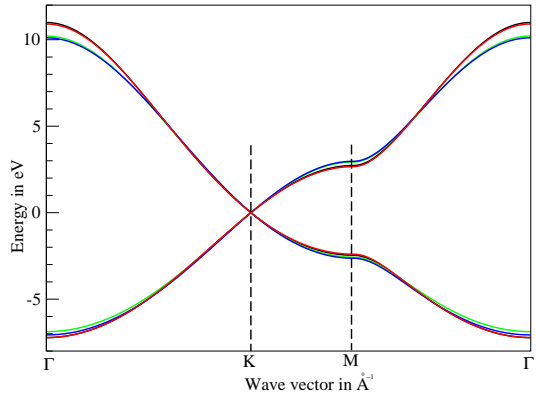}
		\caption{Electronic structure of graphene with first principle result (black curve), first nearest neighbour interactions (green curve), second nearest neighbour interactions (blue curve) and third nearest neighbour interactions (red curve). The parameters are listed in Table II. Here for each curve the parameters are chosen freely.}
	\end{figure}}
\begin{table}[h]
\caption{Tight binding parameters}
\centering
\begin{tabular}{|c|r|r|r|r|r|r|r|}
\hline\hline
Neighbours & $E_2p$(eV) & $\gamma_0$(eV) & $\gamma_1$(eV) & $\gamma_2$(eV) & $s_0$ & $s_1$ & $s_2$  \\
\hline
1st & 0.0& -2.74& &  & 0.065 &   &   \\ 
\hline
2nd & -0.30 & -2.77 & -0.10 &  & 0.095 &  0.003 &   \\  
\hline
3rd & -0.45 & -2.78 & -0.15 & -0.095 & 0.117 & 0.004 & 0.002 \\
\hline 
\end{tabular}
\end{table} 
\subsection{Density of states of graphene}
The density of states of graphene for the above two cases are shown in figures (6) and (7).
\vspace{0.5 cm}
{\begin{figure}[h]
       \centering
       \includegraphics[width=7cm]{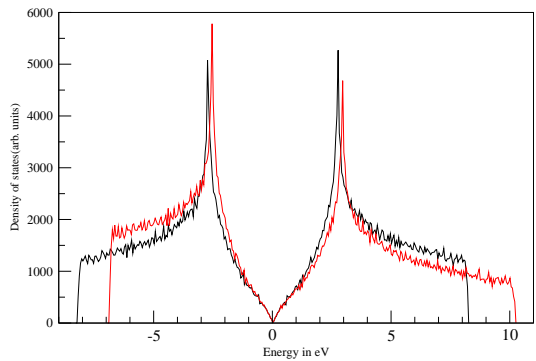}
		\caption{Density of states of graphene with nearest neighbour hopping but zero overlap integral (black curve) and with nearest neighbour hopping and overlap integral (red curve).}
	\end{figure}} 
In fig. (5) we have plotted the density of states of graphene (only for $\pi$ band) for nearest neighbour interaction only without (black curve) and with (red curve) overlap integral correction. It shows that without overlap valence and conduction bands are symmetric but in presence of overlap width of valence band decreases and that of conduction band increases. Since there are two atoms in each unit cell and each carbon atom has one electron in $p_z$ state, the valence band is completely filled and hence, the fermi level lies at the top of the valence band at zero energy which appears at $K$ and $K'$ points of the brillouin zone in energy momentum space. Graphene has zero density of states at fermi energy and over a very small energy range around zero (i.e. $K$ point energy around which energy dispersion is also linear in momentum) the density of states are varying linearly with energy. Due to the flat part of the band near $M$ point of brillouin zone van-Hove singularities are arising in density of states. The positions of the singular points are symmetric in absence of overlap term while the singularity moves slightly towards fermi energy for valence band and goes slightly away from fermi energy for conduction bands with overlap ($s_0$). In figure (6) and (7) the density of states have been plotted as per the bands in fig. (3) and (4). Here we see that though the density of states is linear near fermi energy, it becomes asymmetric due to the presence of second and third nearest neighbour interactions. As expected the density of states in presence of third nearest neighbours is matching well with the density of states due first principle bands when the parameters are chosen freely. In all other cases the band width is changing slightly and positions of van-Hov singularities are changing over a narrow energy range. When compared with ref. (9), the density of states in presence of second nearest neighbours shows a noticable difference. It is observed that when the second nearest neighbours are chosen for good fitting of low energy part of the spectrum, it gives rise to another van-Hov singularity at conduction band edge but that is removed if the parameters are such that it gives good fitting over the whole energy range. Also, that is suppressed in presence of third nearest neighbour interactions.
{\begin{figure}[h]
       \centering
       \includegraphics[width=7cm]{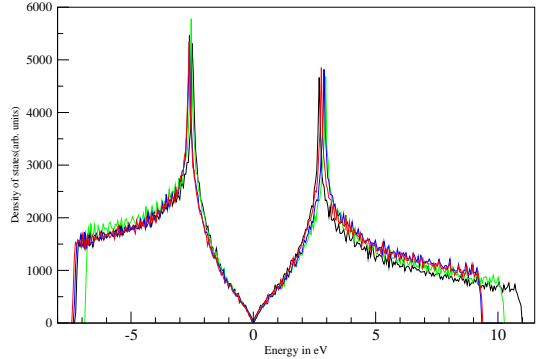}
        	\caption{Density of states of graphene derived from the bands in fig. 3 (parameters are in Table I). First principle result (black curve), first nearest neighbour interactions (green curve), second nearest neighbour interactions (blue curve) and third nearest neighbour interactions (red curve).}
	\end{figure}} 
{\begin{figure}[h]
       \centering
       \includegraphics[width=7cm]{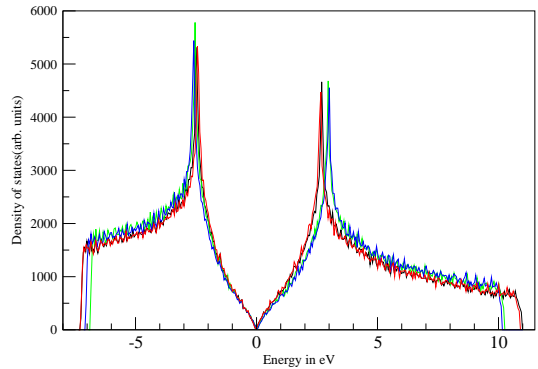}
        	\caption{Density of states of graphene calculated from the bands plotted in fig. 4 (parameters are in Table II). First principle result (black curve), first nearest neighbour interactions (green curve), second nearest neighbour interactions (blue curve) and third nearest neighbour interactions (red curve).}
	\end{figure}} 
\section{Summary and Conconclusion}
To summarise, we have calculated the electronic spectrum and density of states of graphene including upto third nearest neighbour interactions and got a set of tight binding parameters on the physical ground that the absolute values of the parameters should decrease as one moves from first nearest neighbour towards higher distance. This set of parameters can be used to see the effect of intraplane first nearest neighbour overlap integral ($s_0$) and second ($\gamma_1$, $s_1$) and third ($\gamma_2$, $s_2$) nearest neighbour interactions on the band structure of bilayer graphene and graphite.
\section{Acknowledgment}
I am happy to thank Prof. S. G. Mishra for suggesting this work and many useful discussions and thankful to Prof. B. R. Sekhar for his interest in this work and many discussions.
{
\appendix
{\section{Derivation of the electronic dispersions in general and in presence of different neighbours }
In this appendix we present slightly more details of the derivations of dispersion relations in presence of various nearest neighbours. We have the wave function
\be
\Psi_{k}(r)=C_{A}\Psi_{A}^{k}+C_{B}\Psi_{B}^{k}
\ee}
If $H$ is the Hamiltonian and $E(k)$ the eigenvalue then
\be
H\Psi_{k}(r)=E(k)\Psi_{k}(r)
\ee
which gives the secular equation
\be
\left|\begin{array}{cc}
H_{AA} - E(k)S_{AA} & H_{AB}-E(k)S_{AB}       \\
H^*_{AB}-E(k)S^*_{AB} & H_{BB}-E(k)S_{AA}     \\
\end{array}\right|=0
\ee
\bea
\text{where } H_{xy}&=& \int{\Psi^{*k}_{x}(r)H\Psi^{k}_{y}(r)d\vec{r}}\nonumber \\
&=& H^{*}_{yx}\nonumber \\
\text{and }S_{xy}&=&\int{\Psi^{*k}_{x}(r)\Psi^{k}_{y}(r)d\vec{r}} \nonumber \\
&=&S_{yx}, \nonumber
\eea
where $x$, $y$ represent A and B both. Since two sublattices in graphene are equivalent $H_{AA}=H_{BB}$ and $S_{AA}=S_{BB}$. \\
$\therefore$ the general dispersion relation is
\be
E^\pm(k)=\left[(2E_0-E_1)\pm \sqrt{(E_1-2E_0)^2-4E_1E_2}\right]/2E_3,
\ee
where we define $S_{AA}H_{AA}=E_0$, $H_{AB}S_{AB}^*+H^*_{AB}S_{AB}=E_1$, $H_{AA}^2-H_{AB}H^*_{AB}=E_2$ and $S_{AA}^2-S_{AB}S^*_{AB}=E_3$.
\underline{\bfseries{With Nearest Neighbour Approximation:}}
Here we calculate the elements of the determinant in equation (A3) for the nearest neighbour interactions of the atoms.
\bea
H_{AA}&=&1/N\Sigma_A e^{i\vec{k}.(\vec{r}_{A}-\vec{r}_{A})}\left\langle \Phi_{A}(r-r_{A})|H|\Phi_{A}(r-r_{A})\right\rangle \approx E_{2p}. \nonumber \\
S_{AA}&=&\left\langle \Phi_{A}(r-r_{A})|\Phi_{A}(r-r_{A})\right\rangle=1 \text{ (since the wave functions are normalized)}. \nonumber \\
H_{AB}&=&\gamma_{0} \Sigma_B e^{i\vec{k}.(\vec{r}_{B}-\vec{r}_{A})} \nonumber \\
&=&\gamma_0 f(k), \nonumber \\
\text{where }\gamma_0&=&1/N\Sigma_A \left\langle \Phi_{A}(r-r_{A})|H|\Phi_{B}(r-r_{B})\right\rangle \text{ is the hopping integral (-ve) and } \nonumber\\
f(k)&=&e^{ik_{x}a/\sqrt{3}}+2e^{-ik_{x}a/2\sqrt{3}}\cos{k_{y}a/2}. \nonumber
\eea
\bea
\text{Similarly, }S_{AB}&=&s_0f(k), \nonumber \\
\text{where }s_0&=&1/N\Sigma_A \left\langle \Phi_{A}(r-r_{A})\Phi_{B}(r-r_{B})\right\rangle \text{is the overlap integral (+ve).} \nonumber \\
\therefore E_0&=&E_{2p} \nonumber\\
E_1&=&2\gamma_0 s_0|f(k)|^2 \nonumber\\
E_2&=&E_{2p}^2-\gamma_{0}^2|f(k)|^2 \nonumber\\
\text{and }E_3&=&1-s_{0}^2|f(k)|^2 \nonumber \\
\text{Also, } |f(k)|^2&=&g(k) \nonumber \\
&=&1+4\cos^2{(k_ya/2)}+4\cos{(\sqrt{3}k_xa/2)}\cos{(k_ya/2)}\nonumber \\
\implies E^\pm(k)&=&\left[ \left( E_{2p}-s_0\gamma_0g(k)\right)\pm \left( \gamma_0-s_0E_{2p}\right) \sqrt{g(k)}\right] /\left[ 1-s^2_0 g(k)\right].
\eea
\underline{\bfseries{An Alternative Expression:}} From the secular determinant the dispersion can alternatively written as\cite{8}
\be
E^\pm(k)=\left( E_{2p}\pm \gamma_{0}\sqrt{g(k)}\right)/\left( 1\pm s_{0}\sqrt{g(k)}\right).
\ee 
\underline{\bfseries{With Second Nearest Neighbour Approximation }}
In this part the matrix elements $H_{AA}$ and $S_{AA}$ are calculated because only these two elements are affected in presence of next nearest neighbours.
\bea
H_{AA}&=& \int{\Psi^{*k}_{A}(r)H\Psi^{k}_{A}(r)d\vec{r}} \nonumber \\
      &=& E_{2p}+\gamma_{1} u(k), \nonumber \\
\text{where } u(k)&=&2\cos{(k_ya)}+4\cos{(k_xa\sqrt{3})}\cos{(k_ya/2)} \text{ and}  \nonumber\\
\gamma_1&=&1/N\Sigma_{A^{'}} \left\langle \Phi_{A^{'}}(r-r_{A^{'}})|H|\Phi_{A}(r-r_{A})\right\rangle \text{ is the second neighbour hopping integral.}\nonumber \\
\text{Similarly, } S_{AA}&=&1+s_{1}u(k)\nonumber\\
s_1&=&1/N\Sigma_{A^{'}}\left\langle \Phi_{A^{'}}(r-r_{A{'}})|\Phi_{A}(r-r_{A})\right\rangle \text{ is the second neighbour overlap integral.} \nonumber
\eea
Putting all the above elements in the secular determinant we get
\be
E^\pm(k)=\left[ E_{2p}+\gamma_{1}u(k)\mp \gamma_{0}\sqrt{g(k)}\right]/\left[ 1+s_{1}u(k)\mp s_{0}\sqrt{g(k)}\right].
\ee 
\underline{\bfseries{With third nearest neighbor approximation:}}
In presence of third nearest neighbours only the matrix elements $H_{AB}$ and $S_{AB}$ get changed but $H_{AA}$ and $S_{AA}$ remain unchanged. Here
\bea
H_{AB}&=& \int{\Psi^{*k}_{B}(r)H\Psi^{k}_{A}(r)d\vec{r}} \nonumber \\
      &=& \gamma_{0} f(k)+\gamma_{2} v(k), \nonumber \\
\text{where}\gamma_2&=&1/N\Sigma_A \left\langle \Phi_{A}(r-r_{A})|H|\Phi_{B}(r-r_{B})\right\rangle \text{ is the third nearest neighbour hopping integral} \nonumber \\
\text{and }v(k)&=&e^{ik_xa/\sqrt{3}}2\cos{k_ya}+e^{-2ik_xa/\sqrt{3}}. \nonumber \\
\text{Similarly, } S_{AB}&=&s_{0} f(k)+ s_{2} v(k), \nonumber \\
\text{where }s_{2}&=&1/N\Sigma_{A} \left\langle \Phi_{A}(r-r_{A})|\Phi_{B}(r-r_{B})\right\rangle.\nonumber
\eea
$\therefore$ expressions for $E_1$, $E_2$, $E_3$  are
\bea
E_1&=& 2s_0\gamma_0 g(k)+\left( s_0 \gamma_2+\gamma_0 s_2\right) t(k)+2s_2\gamma_2 g(2k) \nonumber \\
E_2&=&\left[ E_{2p}+\gamma_1u(k)\right] ^2-\left[ \gamma^{2}_0g(k)+\gamma_0 \gamma_2 t(k) +\gamma^{2}_2 g(2k)\right]   \nonumber\\
E_3&=&\left[ 1+s_1u(k)\right] ^2-\left[  s^{2}_0g(k)+ s_0 s_2 t(k) + s^{2}_2 g(2k)\right],   \nonumber \\
\text{where } g(2k)&=&1+4\cos^2{(k_ya)}+\cos{(\sqrt{3}k_xa)}\cos{(k_ya)}\nonumber\\
t(k)&=&2\cos{(k_xa\sqrt{3})}+4\cos{(k_ya)}+4\cos{(k_ya/2)}\cos{(k_xa\sqrt{3}/2)} \nonumber \\  
&&+8\cos{(k_ya)}\cos{(k_ya/2)}\cos{(k_xa\sqrt{3}/2)}. \nonumber
\eea
}

\end{document}